\documentclass[useAMS,usenatbib]{mn2e}

\usepackage{graphicx}
\usepackage{txfonts}
\usepackage{epsfig}
\usepackage{verbatim}
\usepackage{multirow}
\usepackage{xspace}

\def\most{MOST\xspace}
\def\gscone{GSC~00154-01871\xspace}
\def\gscseven{GSC~00154-00785\xspace}

\title[New SPB stars in the field of NGC~2244 discovered by \most]{New SPB stars in the field of the young open cluster NGC~2244 discovered by the \most\thanks{Based on data from \most, a Canadian Space Agency mission operated
by Microsat Systems Canada Inc. (formerly the space division of
Dynacon, Inc.) and the University of Toronto Institute for
Aerospace Studies and the University of British Columbia, with
the assistance of the University of Vienna.} photometric satellite.}

\author[D. Gruber et al.] {D. Gruber$^{1,2}$\thanks{E-mail:
dgruber@mpe.mpg.de (DG)}, H. Saio$^{3}$\thanks{E-mail:  saio@astr.tohoku.ac.jp (HS)}, R. Kuschnig$^{1}$, L. Fossati$^{4}$, G. Handler$^{5}$, K. Zwintz$^{1}$, 
\newauthor  W. W. Weiss$^{1}$,  J. M. Matthews$^{6}$, D. B. Guenther$^{7}$, A. F. J. Moffat$^{8}$, S.M. Rucinski$^{9}$, 
\newauthor D. Sasselov$^{10}$\\
$^{1}$University of Vienna, Institute for Astronomy, T\"urkenschanzstrasse 17, A-1180 Vienna, Austria\\
$^{2}$Max Planck Institut f\"ur Extraterrestrische Physik, Giessenbachstrasse, D-85748 Munich, Germany\\
$^{3}$Astronomical Institute, Graduate School of Science, Tohoku University, Sendai, 980-8578, Japan \\
$^{4}$Department of Physics and Astronomy, Open University, Walton Hall, Milton Keynes MK7 6AA, UK\\
$^{5}$Nicolaus Copernicus Astronomical Center, Bartycka 18, 00-716 Warsaw, Poland\\
$^{6}$Department of Physics and Astronomy, University of British Columbia, 6224 Agricultural Road, Vancouver, BC V6T 1Z1, Canada\\
$^{7}$Department of Astronomy and Physics, St. Mary's University, Halifax, NS B3H 3C3, Canada\\
$^{8}$D\'{e}partment de physique, Universit\'e de Montr\'{e}al C.P. 6128, Succursale Centre-Ville, Montr\'{e}al, QC H3C 3J7, Canada\\
$^{9}$Department of Astronomy \& Astrophysics, David Dunlap Observatory, University of  Toronto P.O. Box 360, Richmond Hill, ON L4C 4Y6, Canada \\
$^{10}$Harvard-Smithsonian Center for Astrophysics, 60 Garden Street, Cambridge, MA 02138, USA \\
}

\begin{document}

\date{}

\pagerange{\pageref{firstpage}--\pageref{lastpage}} \pubyear{2002}

\maketitle

\label{firstpage}

\begin{abstract}
During two weeks of nearly continuous optical photometry of the
young open cluster NGC~2244 obtained by the \most satellite, we
discovered two new SPB stars, \gscseven and \gscone.
We present frequency analyses of the MOST light curves of these
stars, which reveal two oscillation frequencies (0.61 and 0.71
c/d) in \gscseven and two (0.40 and 0.51 c/d) in \gscone.
These frequency ranges are consistent with g-modes of $\ell \leq
2$ excited in models of main-sequence or pre-main-sequence (PMS)
stars of masses 4.5 - 5~$M_{\odot}$ and solar composition $(X, Z)
= (0.7, 0.02)$. Published proper motion measurements and radial
velocities are insufficient to establish unambiguously cluster
membership for these two stars.  However, the PMS
models which fit best their eigenspectra have ages consistent with
NGC~2244. If cluster membership can be confirmed, these would be
the first known PMS SPB stars, and would open a new window on
testing asteroseismically the interior structures of PMS stars.
\end{abstract}


\section{Introduction}
Slowly pulsating B (SPB) stars pulsate in high-order, low-degree gravity modes with periods 
of the order of a few days, showing dense frequency spectra, 
low amplitudes and multi-periodicity with beat periods of months 
and even years \citep{wa87, wa91}. The effective temperature range of known SPB stars is 10,000 -
20,000 K; the mass range is 3 - 8 solar masses.  SPB pulsations
are driven by the $\kappa$-mechanism of the
Fe opacity bump at $T\approx 1.5\times10^5$K \citep{gs93,dmp93}. 
Being trapped deep inside the star, SPB g-modes are very promising 
for asteroseismology of massive stars \citep{de07}. Currently, there are at least 50 confirmed SPB stars and 65 
candidate SPB stars in our Galaxy, 60 in the LMC and 11 in the 
SMC \citep{ko06}. 

Since SPB stars have periods of a few days, it can be difficult to
characterise their eigenspectra accurately with data
from ground-based observatories due to aliasing caused by daily
gaps in the time series. Space-based photometry, with long, nearly
uninterrupted time coverage, is essential to perform the best
frequency analyses for asteroseismology.

The \most (Microvariability \& Oscillations of STars) mission, Canada's first space telescope,  
obtains highly precise optical
photometry of bright stars ($V < 10^{\rm{m}}$) nearly continuously over
time spans up to 60 days \citep{most}. Within two years of \most's launch in 2003, the satellite's
capabilities were expanded to include photometry of the Guide
Stars for each Primary Science Target field.  Many variable stars
have been discovered in this large, nearly random, Guide Star
sample ($9 < V < 6$), spread across the HR Diagram.

In January 2008, \most observed the very young open cluster NGC~2244. 
Embedded in the Rosette Nebula, NGC~2244 is located in the vicinity of the Galactic anti-center ($l=206.306$ $b= -2.072$)
at a distance estimated to be in the range of  $1.4 \lesssim \rmn{D[kpc]} \lesssim 1.7$ \citep[see][]{oi81,pe87,he00,ps02,bb09}.


The age of the cluster was first estimated to be $2-4$~Myr \citep[see][and references therein]{ch07}; a more recent estimate
allows a broader range of 0.2 - 6 Myr \citep{bb09}. 
While many of the cluster stars appear to be younger than 3~Myr, about 45\% of the cluster members may be older than 3~Myr, 
indicating an earlier epoch of star formation \citep{bc02}. Of the younger portion of the sample, 5\% are estimated to be significantly younger than 3~Myr, suggesting star formation is ongoing in NGC~2244. 

\gscseven and \gscone were used as Guide Stars for the
NGC~2244 field and were found to be variable. The data analyses
presented in this paper show the stars to be SPB stars similar
to HD 163830, the first SPB star discovered by \most \citep{ae06}
In Sect.\ref{sec:obs}, we give a short description of the
photometry, and in Sect.\ref{sec:datana}, our analysis techniques and results.
Sect.\ref{sec:membership} deals with evidence for the cluster membership of both
stars. Section \ref{sec:models} describes the theoretical models which best fit
the observed pulsations, and we present discussion and our
conclusions in Sect.\ref{sec:conc}.

\section{Observation and data reduction}
\label{sec:obs}

\begin{figure}
\centering
\epsfig{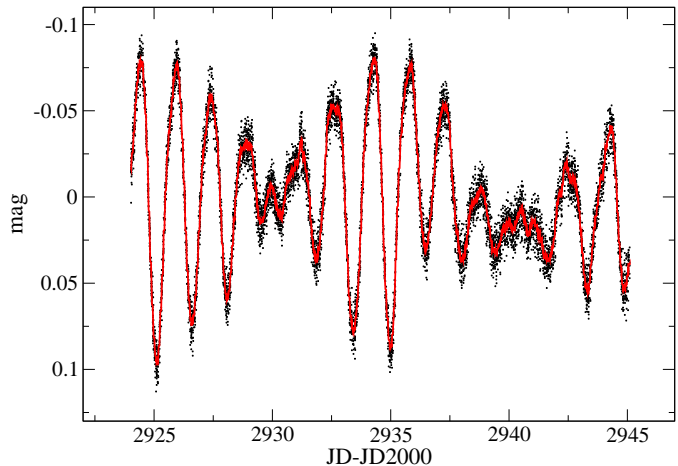}
\caption{
Light curve of \gscseven obtained by MOST photometry for 21 days after subtracting the mean magnitude.
The red solid line shows a boxcar running average with a smoothing length of 50~s. 
}
\label{lc785}
\end{figure}

\begin{figure}
\centering
\epsfig{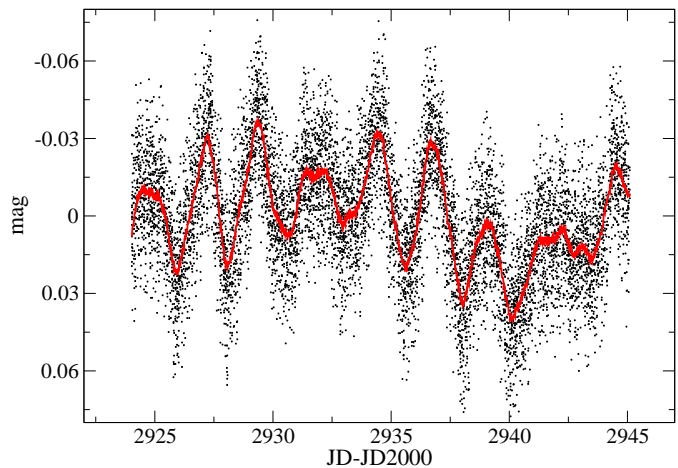}
\caption{
Same as Fig.~\ref{lc785} but for \gscone. 
The red solid line shows a boxcar running average with a smoothing length of 200~s.
}
\label{lc1871}
\end{figure}

The MOST satellite monitored
\gscseven (B9V, V=10.93, $\alpha=6^{h} 33^{m} 35.7^{s}$, $\delta=5^\circ 16' 41''$)
and \gscone (B5V, V=11.30, $\alpha=6^{h} 33^{m} 43.3^{s}$, $\delta=4^\circ 55' 54''$)
as guide stars for science observations of the NGC~2244 cluster field.
Fig.~\ref{lc785} and \ref{lc1871} show the light curves of
\gscseven and \gscone, respectively.
Both light curves clearly show several simultaneous oscillations with
periods near 2 days.

For the NGC~2244 observations 30 subsequent exposures were co-added ('stacked') onboard the satellite with a respective exposure time of 2.03~s. This results in a total sampling period, i.e. the time between consecutive photometric measurements, of 60.9~s. \gscseven was guide star \#35 and \gscone was guide star \# 36 during this observation run.

Stray light due to scattered Earthshine reaches the focal plane
of the \most Science CCD, and this background is modulated with the
orbital frequency of the satellite (14.2 c/d) and some of its
harmonics. (See \citealt{rkfg06} for a discussion of this effect.)
The Guide Star photometric data were folded in phase at the period
of the \most orbit (about 101 min) to treat the orbital modulation
of the stray light background.  Data points most strongly affected
by stray light and other outliers were rejected from the light curve.
To account for the changing nature of the Earth's albedo beneath
the moving satellite, this process was performed with consecutive
segments of the light curve, each four \most orbits long.

We draw attention to a decrease in relative flux of about 0.02 mag
apparent in all the stars in the NGC~2244 field during JD-JD2000 =
2936 - 2940 (and noticeable in Figures \ref{lc785} and \ref{lc1871}).  Although this
artefact does not affect our pulsational frequency analyses below,
we do conservatively neglect any frequencies in the time series
with $f \leq 0.25$ c/d, since we cannot be sure they are intrinsic
to a given star.

\section{Data analysis}
\label{sec:datana}

\subsection{Time series analysis}
For our frequency search, we employed \textit{Period04}\footnote{http://www.univie.ac.at/tops/Period04/} \citep{lenz05}. \textit{Period04} is a code specifically designed to extract multiple
periodic signals in astronomical time series through simultaneous
least squares fitting.

The assessment of the signal-to-noise ratio (S/N) for a given amplitude in general is a complex task. With \textit{Period04}, a mean amplitude is determined in an interval centered on the frequency of interest which is then compared to the amplitude of the peak in question. Obviously, the result depends on the chosen interval which may contain non-noise signal due to intrinsic frequencies, side-lobes, etc. \citet{vaughan05} introduced a method (used by e.g. \citealt{gruber11}) with which one can estimate whether quasi-periodic-pulsations (QPPs) are indeed significant. We applied this method to check which amplitudes exceed the 3.8 $\sigma$ limit commonly used in asteroseismology and find 3 and 2 frequencies passing this requirement for \gscseven and \gscone, respectively.

To determine the uncertainties, we use the well known Rayleigh criterion ($\sigma_f = T^{-1}$, with $T$ being the data set length), which is a conservative estimate of the frequency error as was demonstrated by \citet{ka08}.

\begin{figure}
\centering
\epsfig{file=fig3.eps,width=0.45\textwidth, clip}
\epsfig{file=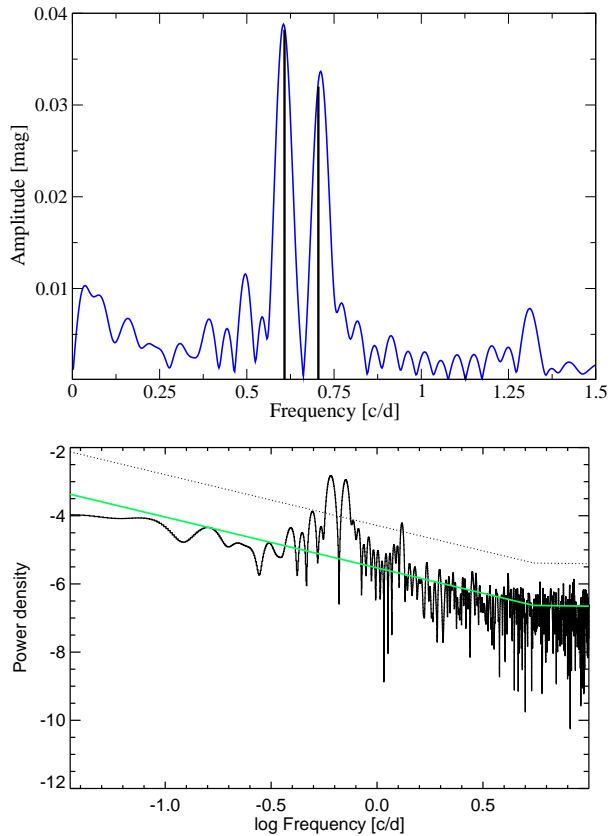, width=0.48\textwidth, clip}
\caption{
\textit{Upper panel:} Region of interest of the amplitude spectrum of \gscseven (blue) 
with the identified frequencies (black). \textit{Lower panel:} Amplitude spectrum in $\log-\log$ space. Shown are the broken power-law fit to the data (green solid line) and the 3.8 $\sigma$ limit (dashed line)
}
\label{fig:gs35ampspec}
\end{figure}
\begin{figure}
\centering
\epsfig{file=fig4.eps,width=0.45\textwidth, clip}
\epsfig{file=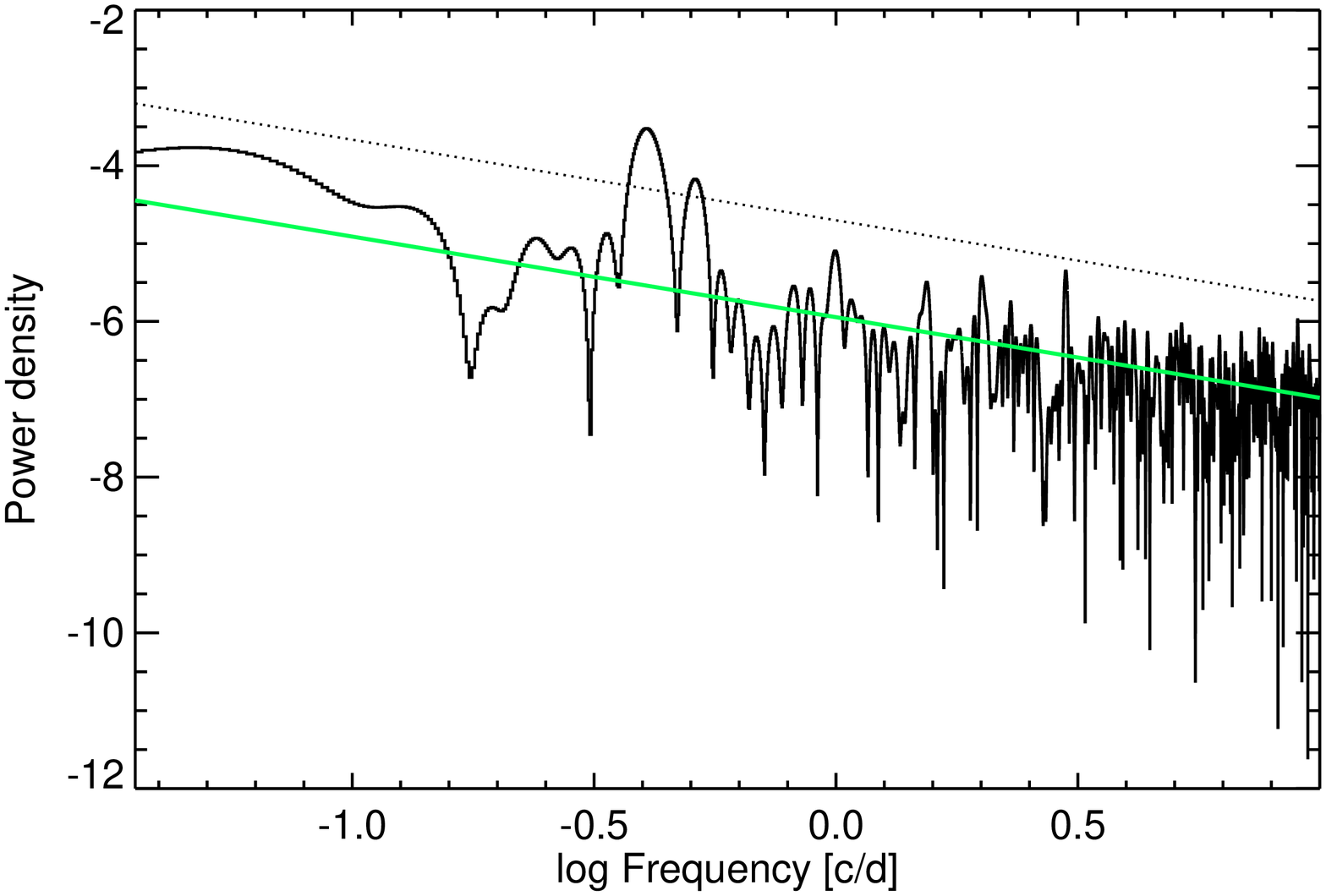,width=0.48\textwidth, clip}
\caption{
\textit{Upper panel:} Region of interest of the amplitude spectrum of \gscseven (blue) 
with the identified frequencies (black). \textit{Lower panel:} Amplitude spectrum in $\log-\log$ space. Shown are the power-law fit to the data (green solid line) and the 3.8 $\sigma$ limit (dashed line)
}
\label{fig:gs36ampspec}
\end{figure}

Table~\ref{tab:freqs} lists the oscillation frequencies and uncertainties, as well as the associated amplitudes that we obtained using the methodology outlined above. Figures~\ref{fig:gs35ampspec} and \ref{fig:gs36ampspec} show the corresponding amplitude and power density spectra.
The number of identified frequencies is small compared to other known SPB stars, as e.g. HD~163830 \citep{ae06}, HD~163868 \citep{walker2005}, and HD~163899 \citep{sa06}. This deficit may be due to a larger noise level of our observations in the frequency domain due to a data set, which is approximately half as long as that for the 3 mentioned SPB stars. 
Additionally, the resulting lower frequency resolution reduces the capability of identifying closely spaced frequencies.

\citet[][and references therein]{de07} argues that amplitude spectra of SPB stars with only a few dominant modes can be explained partially or even entirely by stellar rotation, a concept which was elaborated more recently by \citet{balona11} for $\gamma$Dor stars. If future observations of \gscseven and \gscone do not unravel additional low-amplitude frequencies, they would indeed be candidates for such an explanation. However, if one considers the observed $v \sin i$ of about 8~km/s (see Section \ref{sec:atmparms}), one finds that \gscseven and \gscone would need to have an inclination angle, $i$, of about 4 to 5 degree in order to explain the observed frequencies due to rotation. Assuming a random distribution of $i$ the probability of finding a star with $i<5$ is $P=1-\cos(5)$ which translates to $P \approx 0.4$\%. At this point we cannot rule out this explanation for our observed frequency spectra. But we stress that it is rather unlikely that rotational modulation is the cause for the observed periodicities. In any case, SPB stars with few pulsation frequencies, or few frequencies dominating significantly the frequency spectrum pose a challenge to theory and provide an interesting observational fact in itself.

\begin{table}
\centering
\begin{tabular}{lrccc}
\hline
\multicolumn{4}{c}{\gscseven}\\
 			& \textit{f} [$\rmn{c/d}$] 	& $\sigma$			& amp [mmag]  \\ 
\hline
$f_1$ 		&  0.608 				& $\pm$   	0.048 	&    38.2  		\\
$f_2$ 		&  0.705 				& $\pm$   	0.048 	&    32.0  		\\
$f_3$		& $(f_1+f_2)$ 			& $\pm$	0.048 	&	 6.9	   	\\
\hline
\multicolumn{4}{c}{\gscone}\\
$f_1$ 		&  0.407 				& $\pm$     0.048 	&    17.6  \\
$f_2$ 		&  0.514 				& $\pm$    0.048 	&     8.20  \\
\hline
\end{tabular}
\caption{Frequencies and combination frequencies of \gscseven and \gscone with errors estimated
according to $\sigma=T^{-1}$, and amplitudes (amp).}
\label{tab:freqs}
\end{table}

\subsection{Atmospheric parameters.}
\label{sec:atmparms}


For both stars Str\"omgren \textit{uvby}$\beta$ photometry is available \citep{handinprep}.
Applying the calibrations by \citet{na93} to the data provide
the atmospheric parameters which are listed in Table 2.
\gscone is a spectroscopic binary \citep[see][]{hg06},
therefore the H$\beta$ measurement is highly affected by its companion and thus, not useful to determine $\log g$.

We also analysed high-resolution ($R \cong 48\,000$) spectra of the two stars obtained during 6 - 8 December 2008 with the FEROS spectrograph, installed on the 2.2\,m ESO telescope at La Silla, Chile \citep{ka00}. FEROS is a cross-dispersed, fibre-fed echelle spectrograph. Typical signal-to-noise ratios of the spectra (covering wavelengths from 3800 to 9100 {\AA}) range from 100 to 150. From the FEROS spectra of these stars, provided to us by \citet{niem11}, we obtain similar parameters to those
listed in Table 2, and we confirm that \gscone is indeed a
spectroscopic binary. From the spectral analysis we also report a $v \sin i$ of about 8~km/s for both stars. The corresponding results of the spectroscopic observations will be the subject of a forthcoming paper (Niemczura et al., in preparation).

The average atmospheric parameters (from Str\"omgren photometry
and from spectroscopy), presented in boldface in Table 2, will be
used in the remainder of this paper.

\begin{table*}
\centering
 \begin{minipage}{90mm}
 \caption{Atmospheric parameters for both \gscseven and \gscone determined
from Str\"omgren photometry and spectroscopy.}
\begin{tabular}{llll}
\hline
Star  & $T_{\rm{eff}}$ [K] & $\log g\,[\rm{cm/s^2}]$ & ref. \\
\hline
\multirow{2}{*} {\gscseven}   & $14400 \pm 300$    & $4.0  \pm 0.2 $   & \citet{niem11}\\
                                                              & $14800 \pm 600$     & $4.3 \pm 0.2 $ & \citet{handinprep}\\
                                                              & {\boldmath$14600 \pm 330 $}     & {\boldmath $4.15 \pm 0.15 $} & average\\
\noalign{\smallskip}\hline\noalign{\smallskip}
\multirow{3}{*}{\gscone} &  $13800  \pm 300$  &  $3.8 \pm$ t.b.d   & \citet{niem11} \\
						  &   $13605 \pm 59$ & $3.80 \pm 0.01$ & \citet{hg06} \\
						   &   $14300 \pm 600$ & $4.4 \pm 0.2$ & \citet{handinprep} \\
						   & {\boldmath $13900 \pm 360$} & {\boldmath $3.8 \pm 0.15$} & average\footnote{for the average of $\log g$ the value obtained from photometry was excluded because H$\beta$ is strongly affected by the companion.} \\
\hline
\end{tabular}
\end{minipage}
 \label{tab:atm}
\end{table*}

\section{Field stars or cluster members?}
\label{sec:membership}

\begin{table*}
\caption[ ]{Proper motion and radial velocity of the NGC~2244 open cluster and the two target stars: \gscseven and \gscone. All values, besides the stellar radial velocity, are taken from \citet{kharchenko2004} and \citet{kharchenko2005}.
The radial velocity of \gscone is missing because the star is a spectroscopic binary and we do not have enough spectra to measure a reliable mean radial velocity.}
\label{velocity}
\begin{center}
\begin{tabular}{ccccccccccccc}
\hline
 & RA 	& DEC 	& $\mu_{\alpha}\cos(\delta)$ 	& $\sigma_{\mu_{\alpha}\cos(\delta)}$ & $\mu_{\delta}$ & $\sigma_{\mu_{\delta}}$ 	&$\rmn{V}_{rad}$      & $\sigma_{\rmn{V}_{rad}}$  \\
 &    [J2000]		& [J2000]    		& [mas/yr]  	       			& [mas/yr]			       				& [mas/yr]	& [mas/yr]		  			& [km/s]  & [km/s] \\
\hline											
NGC~2244        & 06 31 55 & +04 56 30 & $-$1.42 & 0.39 & $-$0.32 & 0.37 & 26.16 & 3.37 \\
\gscseven & 06 33 36 & +05 16 41 & $+$1.87 & 1.90 & $-$4.26 & 2.60 & 28.2  & 1  \\
\gscone & 06 33 43 & +04 55 54 & $-$1.16 & 1.78 & $-$3.21 & 1.70 & $--$  & $--$ \\ 
\hline
\end{tabular}
\end{center}
\smallskip

\end{table*}

\subsection{Proper motion study and radial velocity}
To establish membership through a proper motion study, NGC~2244
is not an easy cluster to investigate, due to both its rather
large distance and the similarity between the proper motions of
stars in the cluster and those in the field, as is illustrated
in the atlas of NGC~2244 published by  \citet{kharchenko2005}.
According to this atlas, the cluster member proper motions are
concentrated around a mean of ($-$1.42, $-$0.32)\,mas/yr, compared
to the mean for field stars in the same region of the sky of ($+$0.92, $-$6.55)\,mas/yr with 
a dispersion for the cluster stars of
$\sigma_{\mu_{AC}} = 4.45 \pm 0.15$\,mas/yr \citep{kharchenko2004,ch07}. 

Table~\ref{velocity} shows a comparison between the cluster mean values of
proper motion and radial velocity for NGC~2244 and those of the
two target stars: \gscseven and \gscone. From their
independent proper motion study, \citet{ch07} estimate
membership probability for \gscseven to be 50\%, and for \gscone, 84\%.

For \gscseven, the measurements of both $\mu_\alpha$cos($\delta$)
and $\mu_\delta$ agree with those of the cluster within $2\sigma$. The agreement of
$\mu_\alpha$cos($\delta$) values improves if one considers only
the 30 cluster members listed by \citet{kharchenko2004} with
measurement uncertainties smaller or comparable to those of
\gscseven. In that sub-sample, the dispersion of
$\mu_\alpha$cos($\delta$) is approximately 2.5 mas/yr. The
agreement between the cluster mean values and those of the two
stars is as satisfactory for \gscseven as it is for the most
precisely measured stars in the cluster.

The agreement between the mean cluster radial velocity
\citep{kharchenko2005} and the value given for \gscone
by \citet{niem11} strengthens the case for membership of
this star. However, we do not include this star's radial velocity
in Table~\ref{velocity}, because the star is a spectroscopic binary \citep{hg06}
 and we do not have enough spectra over time to be able
to provide a very accurate mean radial velocity of the binary
system. For \gscone, we arrive at the same conclusion as
for \gscseven, but even more strongly, since the star's value
of $\mu_\alpha$cos($\delta$) agrees with the cluster mean to within
only $1\sigma$.

\subsection{Reddening and CMD analysis}
\label{sec:clusmem}

\begin{figure}
\centering
\epsfig{file=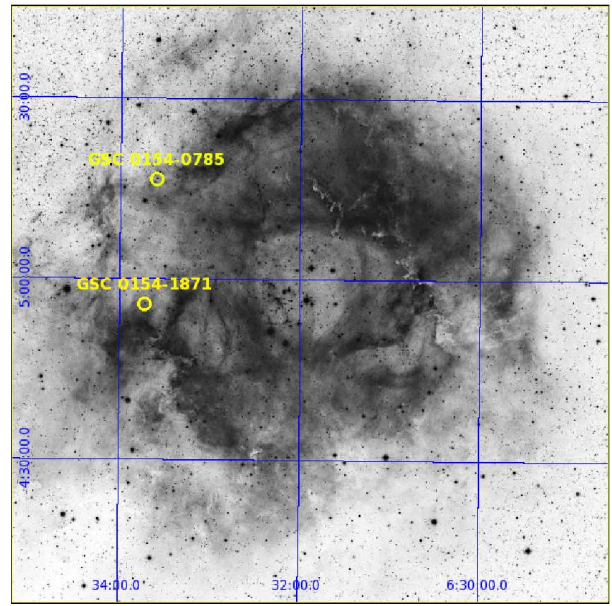, width=0.45\textwidth, clip}
\caption{POSS-I image of NGC~2244 observed at a wavelength of $\sim 0.5 \mu$m. Green circles denote the positions of the two guide stars, \gscseven and \gscone.}
\label{fig:ngc2244}
\end{figure}

In Fig.~\ref{fig:ngc2244} we show the positions of the two stars 
relative to the central region of NGC~2244. 

Applying the calibration by \citet{crawford78} to the Str\"omgren
 colour indices gives $E(b-y)=0.245$ and $M_v=-0.09\pm0.4$ for 
\gscseven and $E(b-y)=0.379$, $M_v=0.15\pm0.4$ for \gscone. 
The considerably lower reddening of \gscseven is 
understandable given its position in the outer parts of the region of NGC~2244.
With $V$ magnitudes of 10.93 and 11.30 respectively \citep{handinprep}, we obtain distances of 
$980\pm400$ and $800\pm350$ pc for \gscseven and \gscone, respectively.

The distance of NGC~2244 is estimated to fall in the range of
1.4 - 1.7 kpc \citep[see][and references therein]{bb09}.
Our Str\"omgren results put the two stars at roughly half the
cluster distance and thus argue against cluster membership.
However, we point out that the distances we derive for the stars
are within $1\sigma$ of the lowest estimate of the cluster distance.

Words of warning:  If either star is surrounded by circumstellar
matter, this would redden the stars, making the
\citet{crawford78} calibration routine invalid.  However, neither of
the two stars shows emission lines in its spectrum \citep{niem11}. \gscone is a spectroscopic binary and therefore its
absolute magnitude derived from $uvby$ photometry cannot be trusted.
For B stars, the latter is calibrated via H$\beta$ whose value is
affected by the presence of a companion.

The calibration routine implicitly assumes that the two
stars are main-sequence (MS) stars. If these stars are in their
pre-main sequence (PMS) stage, we argue that the Crawford (1978)
routine can be safely applied because (a) there is no evidence for
circumstellar emission, and (b) the colour excesses from the
$uvby$ photometry are not anomalous.

\begin{figure}
\centering
\epsfig{file=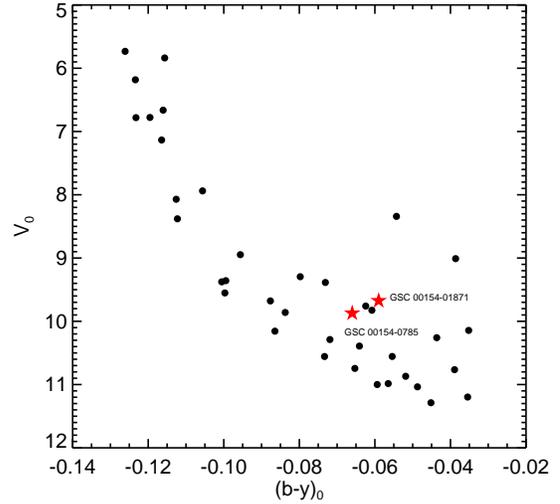,width=0.5\textwidth, clip}
\caption{
Zoom into the dereddened colour-magnitude diagram (CMD) for cluster members of NGC~2244. Positions of \gscseven and \gscone are superimposed (red stars).
}
\label{fig:cmd}
\end{figure}

Figure \ref{fig:cmd} is a dereddened colour-magnitude diagram (CMD) containing
stars from the samples of \citet{heiser77, pe89,kaltcheva99, he00} and \citet{handinprep}.
These have a membership probability $> 85$\% (cross-checked against
\citealt{marschall82, kharchenko2005, baumgardt00} and \citealt{ch07}). The dereddened colour index $(b-y)_0$  and $V_0$  were obtained using the \citet{crawford78} calibration
as described above. The unsuspicious position of \gscseven and \gscone in the CMD as well as their location in the field of NGC~2244 supports cluster membership of the two stars.

\subsection{Conclusion of this section}
Neither the cluster membership determinations nor the distance
estimates are iron-clad, and they suffer from qualifiers and
caveats. Although, membership of \gscseven and \gscone
is possible, we are unable to confirm it. GAIA parallaxes will settle this issue.

\section{Models}
\label{sec:models}
\subsection{Evolutionary models}

\subsubsection*{\gscseven}

The effective temperature and surface gravity derived in Sect.~\ref{sec:atmparms}
are plotted in the bottom panel of Fig.~\ref{fig:hrdgte} with main-sequence (MS)
evolutionary tracks\footnote{Evolutionary models were computed by a standard Henyey type code.} (solid lines) and pre-main-sequence (PMS) tracks
(dashed lines).
This diagram indicates that \gscseven has a mass
of $ 4.5 \pm 0.5\,M_{\odot}$ no matter whether it is an MS or PMS
star.

We can estimate the age of the star by comparing its position on the $\log g - \log T_{\rm eff}$ diagram with the evolutionary tracks. If the star is assumed to be a MS star, its age is $20-50$~Myr.

Assuming it to be a \textit{field star}, thus using
the absolute magnitude derived from Str\"omgren photometry (Sect. \ref{sec:clusmem}),
and applying a bolometric correction of $-1.14$ \citep{fl96},
we find $M_{\rm bol}=-1.23\pm0.4$
which corresponds to a luminosity of
$\log L/L_\odot = 2.42 \pm 0.16$. 
However, if \gscseven is a \textit{cluster member}, adopting a distance modulus of $(m-M)_0=11.1$ for~NGC~2244 \citep[see e.g.][and references therein]{ps02} we find $M_{\rm bol}=-2.27$, corresponding to a luminosity of $\log L/L_\odot = 2.8\pm0.25$.

\subsubsection*{\gscone}
\begin{figure}
\epsfig{file=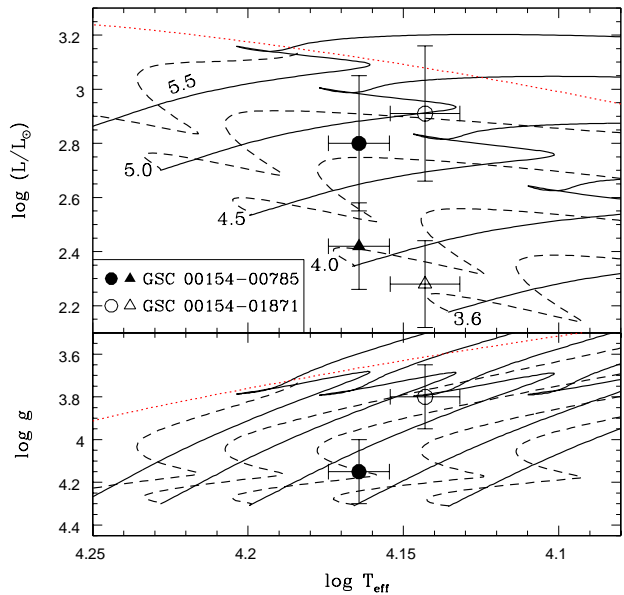,width=0.5\textwidth}
\caption{
Estimated parameters for \gscseven (filled symbols) 
and \gscone (open symbols) are compared with evolutionary models 
in the HR diagram (top panel) and in the $\log g-\log T_{\rm eff}$ 
diagram. The two positions for each star in the top panel correspond 
to the two kinds of estimates for the luminosity discussed in the text: circles denote the luminosity assuming the stars to be cluster members, triangles denote the field star assumption. 
Dashed lines are PMS evolution tracks, while solid lines are  
MS and post-MS evolution tracks for each stellar mass indicated 
(in solar units) along each track in the top panel. 
Dotted lines indicate the approximate locus of the birth line. 
}
\label{fig:hrdgte}
\end{figure}

The position of \gscone in Fig.~\ref{fig:hrdgte} is also based on the parameters derived in Sect.~\ref{sec:atmparms}.
The location of this star in
Fig.~\ref{fig:hrdgte} suggests that this star has a mass of $5.0 \pm 0.5
\,M_{\odot}$, whether it is an MS or PMS star.  If the star is
assumed to be an MS star, its age is $70-80$~Myr.

We estimate the luminosity as above for \gscseven.
With a bolometric correction of
$-1.02$ \citep{fl96}, we find for \gscone the following values: $M_{\rm bol}=-0.87$ and 
a luminosity of $\log L/L_\odot= 2.28\pm0.16$. 
On the other hand, \gscone has a stronger case for \textit{cluster} membership. If we adopt the same distance modulus
as for \gscseven above and after applying a bolometric correction of $-1.02$ \citep{fl96} we find a bolometric magnitude of $M_{\rm bol}=-2.45$ and a corresponding luminosity of
$\log L/L_\odot= 2.91\pm0.25$.

If we assume that \gscone is a field star, the evolutionary models
that best fit this star in a $\log L - \log T_{\rm eff}$ plot conflict with what
is expected from the $\log g - \log T_{\rm eff}$ diagram shown in Fig.~\ref{fig:hrdgte}. On the
other hand, the stellar masses derived from the evolutionary models are consistent in both diagrams
if the star is assumed to be a cluster member. Additionally, if the star is in
the field, its luminosity would be only marginally consistent with
models of g-mode excitation.  Both points are circumstantial
evidence to support the membership of \gscone in NGC~2244.

\subsubsection*{Implications: MS or PMS?}

If these stars are on the MS, the estimated ages are much larger
than the age range for the open cluster NGC~2244 \citep[$0.2 - 6$~Myr][]{bb09}. This argues against a cluster membership if they are MS stars.
However, if the stars are still in their PMS
stage, the situation is different. The dashed lines in Fig.~\ref{fig:hrdgte} show
PMS evolutionary tracks. The initial model for each PMS track has
been obtained by accreting mass starting from a 1~$M_{\odot}$
model on the Hayashi track. As seen in Fig.~\ref{fig:hrdgte}, both stars can be
identified as PMS stars with masses near $4.5 \,M_{\odot}$. The ages
are estimated for \gscone and \gscseven to be about 0.2~Myr and 0.4~Myr, respectively, after termination of mass accretion.
If the stars are indeed cluster members, then their location in
the HRD implies that they must be PMS stars.  In the following,
pulsational analyses, we consider the g-mode pulsation spectra in
both MS and PMS models. (A detailed discussion about g-modes in PMS stars can be found in Appendix~\ref{appendix}.)

\subsection{Pulsation models}

\begin{figure*}
\epsfig{file=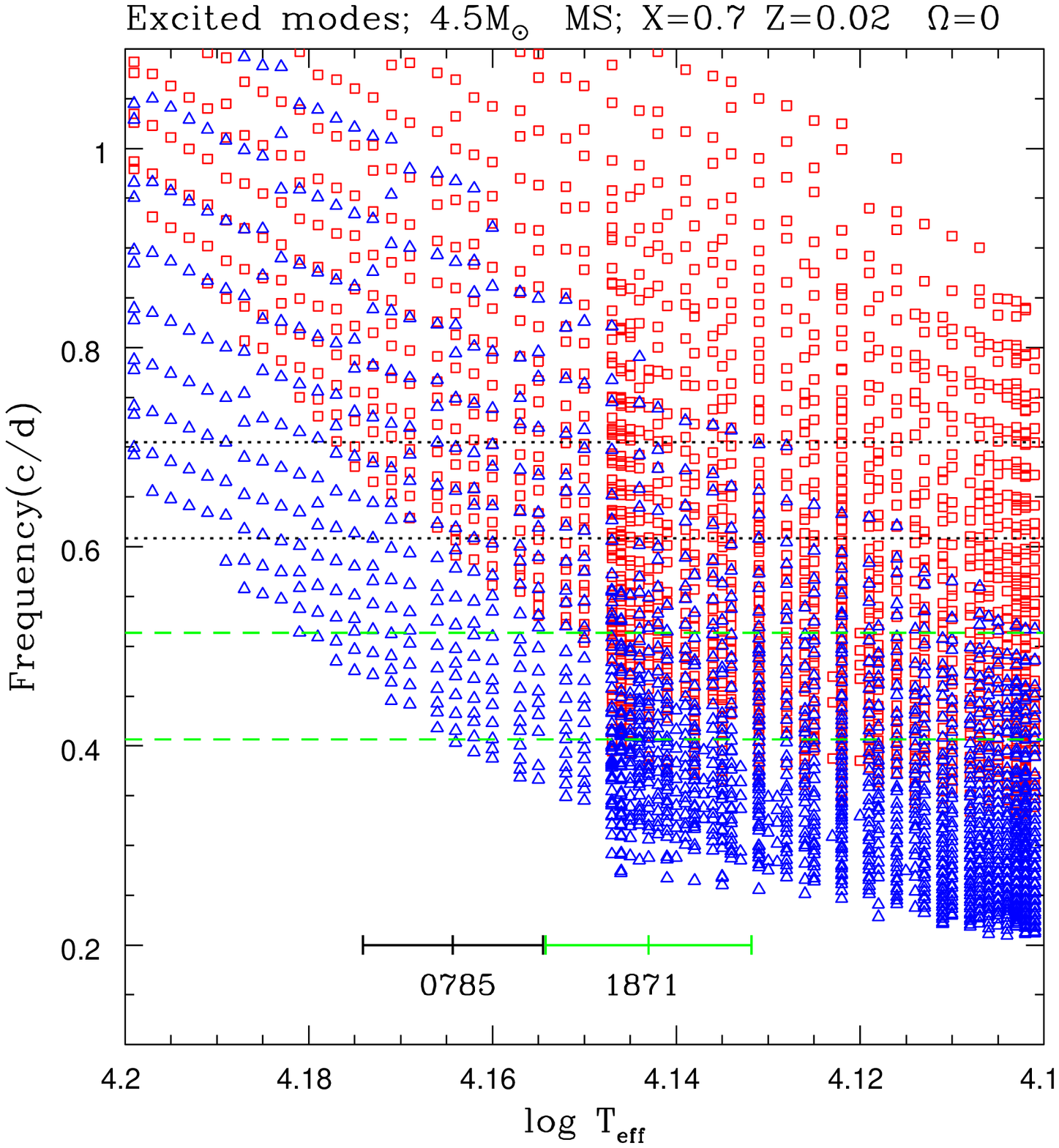,width=0.49\textwidth}
\epsfig{file=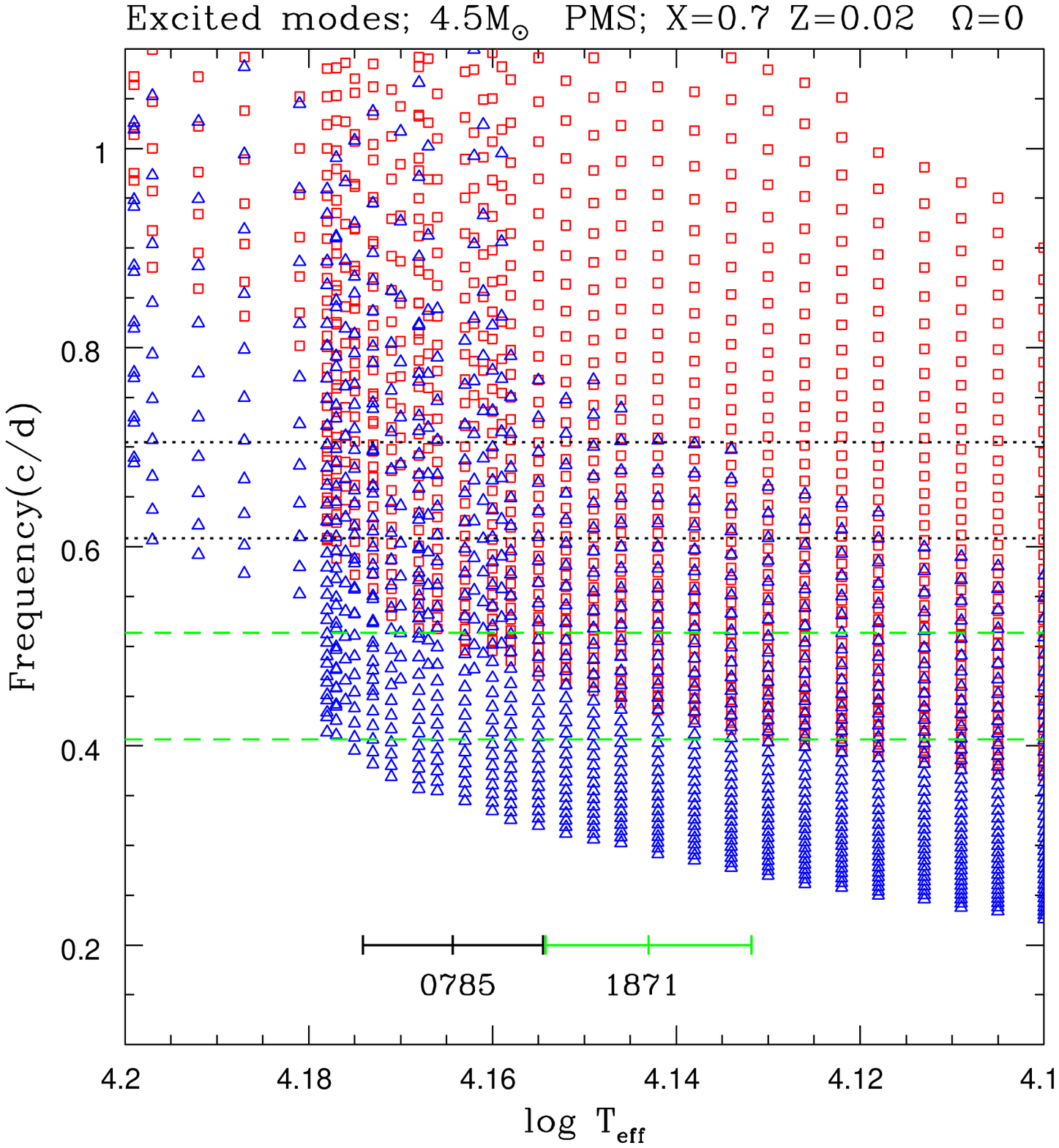,width=0.49\textwidth}
\caption{Frequencies of excited high-order, low-degree $\ell=1$ (blue triangles) and $\ell=2$ (red squares) g-modes
versus effective temperature for the $4.5\,M_\odot$ evolution
models in the MS (left panel) and in the PMS 
(right panel) phases.
Horizontal dotted and dashed lines indicate frequencies detected in 
\gscseven and \gscone, respectively. 
The cross and the horizontal line in the lower part of each diagram
indicates the estimated effective temperatures with error bars for 
\gscseven and \gscone.   
}
\label{fig:nute}
\end{figure*}

Fig.~\ref{fig:nute} shows oscillation frequencies of excited 
low-degree $\ell=1$ and $\ell=2$ modes along the evolutionary tracks 
of non-rotating 
$4.5\,M_\odot$ models in MS (left panel) and
in PMS (right panel) stages. 
We have adopted a standard chemical composition 
($X = 0.7$, $Z = 0.02$) with OP opacity tables \citep{bad05}. 
In a casual inspection of Fig.~\ref{fig:nute}, no significant
differences in the range of excited g-mode frequencies are evident
between MS and PMS models. Both types of models with similar
effective temperatures excite g-modes with similar periods,
because the excitation occurs in outer layers of the star. Before
a convective core appears in the PMS stage (around the peak in
luminosity), the amplitudes of g-modes in the deep interior are
much larger than in MS models, although the number of nodes (a
few tens) are not large enough to significantly dissipate g-mode
oscillations as in the post-main-sequence models. The g-mode spectra for PMS
models are simpler than those for MS models because of the absence
of mode crossings associated with the gradient in mean molecular
weight found after the ZAMS.

The range of excited frequencies shifts downward as the effective
temperature decreases, and for a given effective temperature the
frequency range is slightly smaller for PMS models.  Those trends
are due mainly to the difference in stellar radii of the models;
the frequency of a g-mode is lower for a star of larger radius.
Fig.~\ref{fig:nute} shows that, for both MS and PMS models, the frequency
range of excited $\ell =$ 1 and 2 modes covers well the ranges of
oscillation frequencies detected in \gscone (dashed lines)
and \gscseven (dotted lines).

\begin{figure}
\epsfig{file=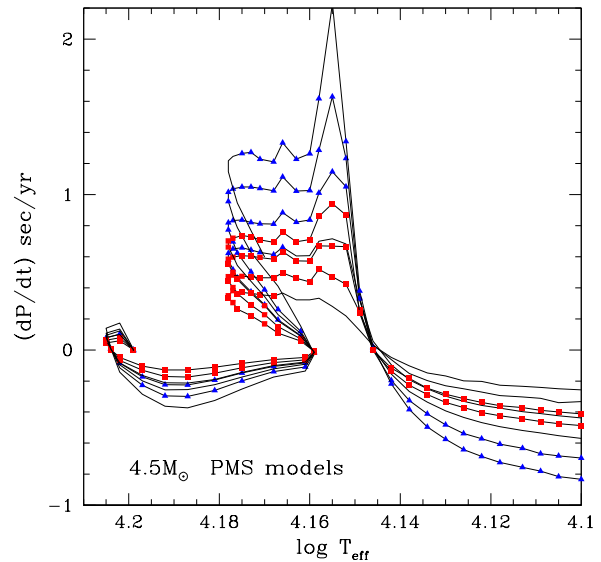,width=0.49\textwidth}
\caption{
The rates of period change for some g-modes in the $4.5\,M_\odot$ model 
during the PMS evolution.
Plotted modes are $g_{15},~g_{20},~g_{25}$ and $g_{30}$ for
$\ell = 1$ and 2. (Longer period modes tend to have larger $|dP/dt|$.)
Filled triangles and squares indicate
modes for $\ell=1$ and 2, respectively to be excited.
}
\label{fig:dfdt}
\end{figure}

If \gscseven and/or \gscone are PMS stars, 
evolutionary
period (frequency) change is expected to be more rapid than in the
MS stage. Fig.~\ref{fig:dfdt} shows the rates of period change $dP/dt$ for
selected g-modes in the 4.5~$M_{\odot}$ model during PMS evolution.

At the effective temperature of \gscseven (log $T_{\rm eff}
= 4.164$), the periods are expected to increase with time,
while at the effective temperature of \gscone (log $T_{\rm
eff} = 4.143$), they decrease.  The period decrease is caused
by the PMS contraction, while the increase is caused by
the fact that Brunt-V\"ais\"al\"a frequency decreases deep in the
star with the development of a convective core which first appears
at log $T_{\rm eff} \simeq 4.15$.  

If \gscseven and \gscone are PMS stars, Fig.~\ref{fig:dfdt} indicates that the period change rates is $\sim 1$ sec/yr and $\sim -0.2$ sec/yr, respectively. Such period changes might be detectable by an (O-C) diagram based on long baseline observations.

\section{Conclusions}
\label{sec:conc}

\most has revealed the SPB nature of stars (\gscseven and \gscone) in the same part of the sky as the very young open
cluster NGC~2244.  The main pulsation periods of these multi-periodic
SPB stars are near 2 days.  The ranges of pulsation frequencies
of the stars is consistent with nonradial g-modes low degree ($\ell = 1$ and $\ell = 2$)
 excited by the kappa mechanism at the Fe bump.

The effective temperatures and surface gravities we derive from
Str\"omgren indices and spectroscopy indicate the two stars have
masses of about 4.5 - 5.0~$M_{\odot}$.  If \gscseven and \gscone are MS stars, our estimates for their ages
are $20-50$~Myr  and $70-80$~Myr, respectively, much older than
the cluster NGC~2244 (0.2 - 6~Myr). If the two stars are still
in the PMS stage, they would be a few tenths of Myr
after the end of the protostellar accretion phase.  Therefore, if
these stars are proven members of NGC~2244, then almost certainly
they are PMS stars.  Existing proper motion and radial velocity
properties of the two pulsators cannot establish reliably
their membership in the NGC~2244 cluster. GAIA parallaxes
and proper motions should eventually remove this uncertainty.

If the membership to NGC~2244 can be indeed confirmed, \gscseven and \gscone would be the first known PMS SPB pulsators.  This
possibility has alerted us to begin a more concerted search with
\most and other space photometry missions for other SPB pulsators
among PMS cluster and field stars.


\section*{Acknowledgments}
The authors are grateful to Ewa Niemczura for useful discussion and the analysis of the Feros spectra. We dedicate this work to the memory of our dear friend and colleague Dr. Piet Reegen, who will be greatly missed by us all.
Astronomy research at the Open University is supported by an STFC rolling grant (L.F.).
GH is partially supported by the
Austrian Fonds zur F\"orderung der wissenschaftlichen Forschung under
grant P20526-N16 and WW by P22691-N16.
KZ is recipient of an APART fellowship of the Austrian Academy of Sciences at the Institute of Astronomy of the University Vienna.
J.M.M., A.F.J.M, and S.M.R. are funded by NSERC (Canada) and some
of R.K.'s contributions were partially funded by the Canadian
Space Agency.

\appendix
\section{Instability range for  g-modes in PMS stars}
\label{appendix}
Since PMS evolutionary tracks of intermediate masses 
($3\la M/M_\odot\la 6$) enter into a g-mode instability region (i.e., the SPB instability region) 
caused by the Fe opacity bump at $T\approx 1.5\times10^5$~K,
certain g-modes can theoretically be excited in some intermediate PMS stars. 
We have examined the instabilities of nonradial g-mode oscillations of $\ell=1$ and 2 for PMS models 
with masses ranging across 2.7~$M_\odot$ to 6~$M_\odot$.
The initial models for the evolution of each mass were obtained by accreting mass at a rate of
$10^{-5}\,M_\odot$yr$^{-1}$ starting from a fully convective 1~$M_\odot$ model.
Those initial models are roughly located on the birth-line of PMS stars \citep{ps93}.

\begin{figure}
\epsfig{file=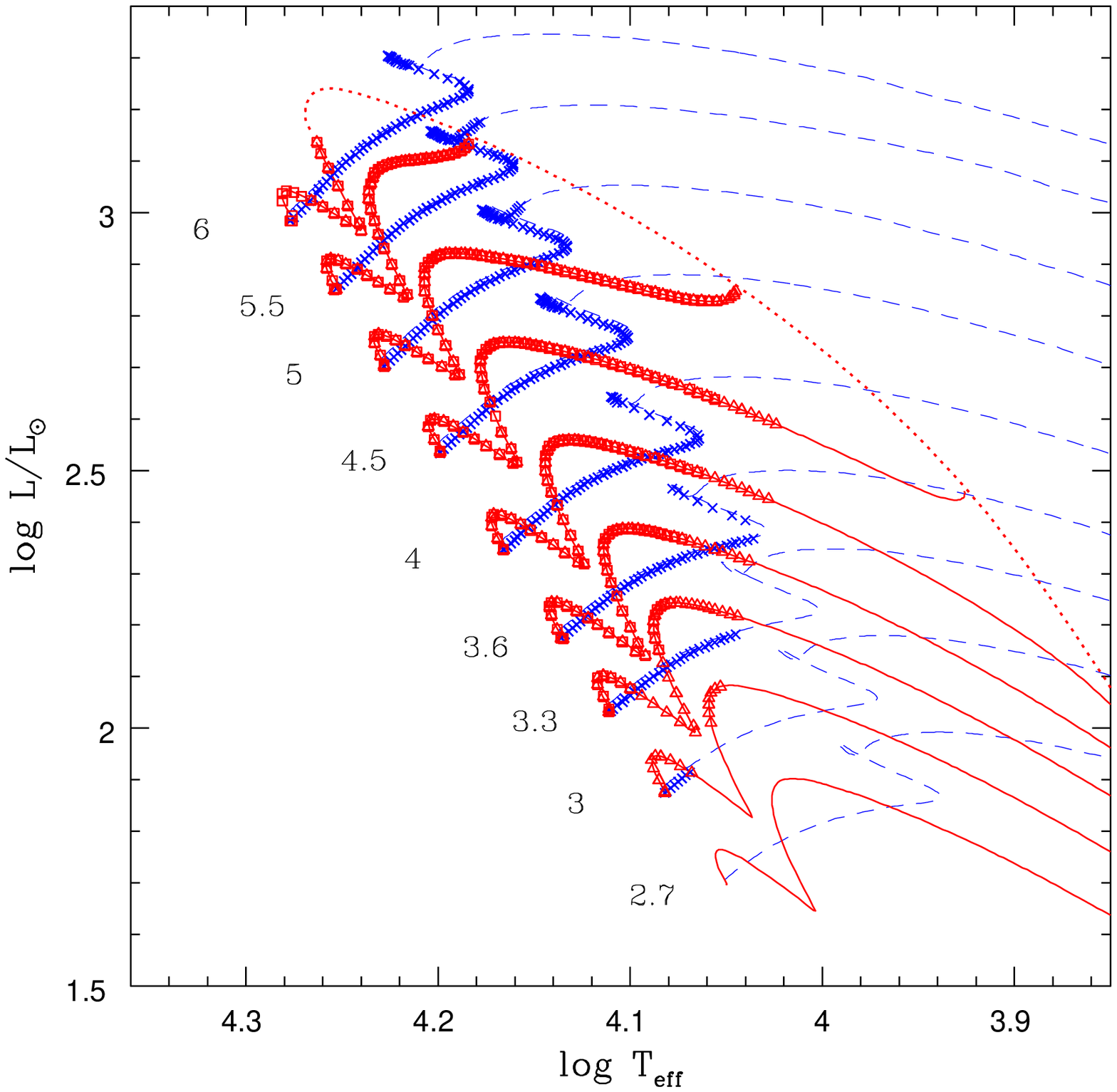,width=0.49\textwidth}
\caption{Models in which g-modes are excited are indicated by symbols
($\triangle$ and $\square$ for $\ell=1$ and 2 modes, respectively for PMS
models,
and $\times$ for both $\ell$s for main-sequence models ) along
evolutionary tracks with various masses.
Red solid lines indicate PMS models, while blue dashed lines main-sequence
and post-main sequence models.
Dotted line indicates evolutionary tracks during mass-accreting
proto-stellar evolution.
}
\label{prem_inst}
\end{figure}

The stability results are shown in Fig.~\ref{prem_inst}, where red triangles
and squares indicate the positions of models in which g-modes are
excited. Red lines indicate the evolutionary tracks of PMS stars,
while blue lines are main-sequence and post-main-sequence tracks.
Main-sequence models unstable to g-modes (corresponding to what
would be currently recognised SPB stars) are highlighted by blue
crosses.

Because the Brunt-V\"ais\"al\"a frequency becomes very high in the core
of a post-main-sequence star, g-modes are damped due to dissipation 
unless a shell convection zone appears, as in massive stars \citep{sa06,ga09}. 
This is why the instability range for prototype SPB stars 
is confined to the MS band \citep{gs93,dmp93}.
In PMS stars, however, g-modes survive even in the phase without core convection, 
because the Brunt-V\"ail\"al\"a frequency in the interior is not as high as 
in a post-main-sequence star. The spatial wavelength of the eigenfunction
is therefore too long for the dissipation to be important.
(The number of nodes of excited g-modes is well below 100.) As a result, the g-mode
instability range for PMS stars (red symbols in Fig.~\ref{prem_inst}) is broader
than that for MS SPB stars (blue symbols). The models
predict PMS SPB stars which overlap with and extend beyond the
main sequence band in the HR Diagram.

\end{document}